# Avian magnetoreception model realized by coupling magnetite-based mechanism with radical-pair-based mechanism

Yan Lu[1] and Tao Song[1,2]*

[1]Beijing Key Laboratory of Bioelectromagnetism, Institute of Electrical Engineering, Chinese Academy of Sciences, P. O. Box 2703, Beijing 100190, P. R. China

[2]France-China Bio-Mineralization and Nano-Structures Laboratory (BioMNSL), P. O. Box 2703, Beijing 100190, P. R. China

*Corresponding author (songtao@mail.iee.ac.cn)

ABSTRACT

Many animal species were verified to use geomagnetic field for their navigation, but the biophysical mechanism of magnetoreception has remained enigmatic. This paper presents a special biophysical model that consists of magnetite-based and radical-pair-based mechanisms for avian magnetoreception. The amplitude of the resultant magnetic field around the magnetic particles corresponds to the geomagnetic field direction and affects the yield of singlet/triplet state products in the radical-pair reactions. Therefore, in the proposed model, the singlet/triplet state product yields are related to the geomagnetic field information for orientational detection. The resultant magnetic fields corresponding to two materials with different magnetic properties were analyzed under different geomagnetic field directions. The results showed that ferromagnetic particles in organisms can provide more significant changes in singlet state products than superparamagnetic particles, and the period of variation for the singlet state products with an included angle in the geomagnetic field is approximately 180° when the magnetic particles are ferromagnetic materials, consistent with the experimental results obtained from avian magnetic compass. Further, the calculated results of the singlet state products in a reception plane showed that the proposed model can explain the avian magnetoreception mechanism with an inclination compass.

## Ⅰ. INTRODUCTION

Various animals use geomagnetic field for their navigation. The use of magnetic compass by migratory birds was first described in European robins [1]. In the following years, it has been further demonstrated in more than 50 species [2]. Recently, two most convincing biophysical mechanisms of magnetoreception in birds have been proposed: magnetite-based mechanism [3–5] and radical pair-based mechanism [6–8].

The magnetite-based mechanism is established on the discoveries that magnetic particles are widely found in bacteria and higher organisms [9, 10]. These magnetic particles were suggested to be involved in magnetoreception, verified by various biophysical methods. Some magnetite-based models have been presented and theoretically evaluated in efforts to explain magnetoreception [11–15].

The model proposed by Kirschvink et al. [11, 12] was based on the presumption that the magnetic particles in organisms are single-domain (SD) magnetites. The authors believed that external magnetic field exerts a magnetic torque on the particle, which rotates the particle to open or close the ion channel and produces nerve signals. The model presented by Davila et al. [13] was based on superparamagnetic (SP) nanoparticles. They assumed that SP nanoparticles can be magnetized, and the local magnetic field in the cell is amplified by orders of magnitude under external magnetic field. Thus, magnetic particles experience an attractive or repulsive force to induce their displacement, which can induce primary receptor potential via strain-sensitive membrane channels to create nerve signals. Fleissner et al. [14], and Solov'yov et al. [15] analyzed the magnetoreception process in birds using two types of ferromagnetic magnetite particles based on the model of Davila et al. [13]. All the models mentioned above indicated that nerve signals are transmitted through the ophthalmic branch of the trigeminal nerve to the brain, and magnetic fields are used as compass or component of a navigational "map" for the navigation of birds [16, 17].

However, these theories cannot explain the light dependence of magnetic orientation confirmed by several behavioral experiments [18–20]. Further, Zapka et al. [21] reported that European robins with bilateral section in the ophthalmic branch of the trigeminal nerve can still use their magnetic compass for orientation, in conflict with the magnetite-based mechanism because the mechanism requires trigeminal nerve signal transmission.

On the other hand, the radical-pair-based mechanism is established on the proposition that weak magnetic fields can affect free-radical recombination reactions [22–24]. The singlet/triplet radical-pair product can act as the receptor of the external magnetic field, and migratory birds can use it to sense changes in the geomagnetic field. The magnetosensitivity of the radical-pair reactions has to be anisotropic to detect the direction information of the geomagnetic field using the radical-pair mechanism. Theoretically, the molecules that provide the radical pairs can possess the required properties to produce anisotropic effect [25, 26]. Thus, Ritz et al. [27] proposed a special structure of radical pairs called "reference-probe" design, where the internal magnetic field of the "reference" radical should be much higher than the external field, but the "probe" radical has very small or no internal magnetic field. Lau et al. [28] suggested that the molecules that play host to the magnetically sensitive radical-pair intermediates must be immobilized and rotationally ordered within receptor cells. Then, a partially rotational disorder can cause anisotropic responses of differently oriented radical pairs within the same cell. Although anisotropic radical-pair magnetic field effect has been observed in several solutions or liquid crystal experiments [29–34], no direct evidence in terms of microcosmic mechanism has been obtained until now.

In addition, some research works observed radical pairs subjected to magnetic fields produced by iron-containing particles and structures [35–39]. Binhi et al. [35] speculated that the rate of intracellular free-radical biochemical reactions will be altered by stray fields produced in different orientations by the intracellular magnet, thought of as an indirect torque-transduction pathway for magnetoreception by Winklhofer et al. [40], who believed that the model can explain the conflicting results reported by Zapka [21]. To achieve maximum angular sensitivity, the Binhi compass model requires that the intracellular magnet should rotate, and the magnet should be coupled to a soft elastic matrix. Cohen et al. [37, 38] demonstrated how magnetic nanostructures can catalyze intersystem crossings in molecular radical pairs. The radical pair was assumed to be randomly distributed and randomly oriented on the surface of the magnetic nanostructure. Cohen et al. modeled the nanostructure as a uniformly magnetized sphere to calculate the magnetic field gradient at the surface of the sphere. The effect of the geomagnetic field on the magnetic field gradient near the magnetic nanostructures was not considered. Cai [39] demonstrated how to optimize the design of a chemical compass using a much better directional sensitivity by simply employing a gradient field created in the vicinity of a hard ferromagnetic nanostructure. He assigned directly the value of the gradient field to analyze the magnetic field sensitivity of the chemical compass without considering the real structure of the nanoparticles in bird.

In this paper, the equivalent model of magnetic particles was established based on the real nanoparticle structure in birds. A special combination of magnetite-based and radical-pair-based mechanisms was analyzed to evaluate the justifiability of the model. In contrast to existing models, the amplitude of the resultant magnetic field around the magnetic particles (but not the magnetic field gradient) was adopted to represent the geomagnetic field direction. Based on the amplitude variation of the resultant magnetic field, the singlet state product yield of the reactions was calculated using the radical-pair mechanism. Thus, product yield is related to the geomagnetic field information.

## Ⅱ. COMBINED MECHANISM

The subtle iron-containing structure in the skin of the upper beak of birds has three subcellular components containing iron: chains of maghemite crystals, magnetite clusters, and iron-coated vesicle [14]. The equivalent model of the magnetic particles in birds was based on this structure. The magnetic field induced by the magnetic particles and the geomagnetic field produce a magnetic resultant field, whose amplitude depends on the spatial location and the direction of the geomagnetic field. Hence, the directional change in the geomagnetic field can be represented by the resultant magnetic field amplitude.

According to the radical-pair mechanism, the radical pair $D^{+}$+$A^{-}$ is created by an electron transferred from donor molecule D to acceptor molecule A. The spin directions of radical pair $D^{+}$+$A^{-}$ have an anti-parallel alignment, and the spin state is called singlet state; when the alignment is parallel, it is called a "triplet state." The radical pair originates from a singlet state; its singlet and triplet states are interconverted by hyperfine interaction, and the process is affected by the applied magnetic field. Therefore, the product yield of a singlet/triplet state varies with the change in the resultant magnetic field arising from the change in the geomagnetic field direction. The radical-pair product yield is thus related to the geomagnetic field information. Consequently, alteration in the product yield will change the number of neurotransmitters and results in an increase or decrease of the signal in the nerve cell that receives the neurotransmitters for navigation. To explain the transmission process, cryptochrome was suggested as a potential primary magnetoreceptor to receive the information of the change in the radical-pair product yields [7].

The proposed mechanism can well overcome the separate disadvantages of the magnetite-based and radical-pair-based mechanisms. Dealing with the role of the magnetic particles in changing the direction information of the geomagnetic field to amplitude of the resultant magnetic field is the core of the proposed model. Past related research has not considered this option. Furthermore, because the equivalent model of magnetic particles was established based on real bird nanoparticle structures, if more convictive experimental evidence for anisotropic interaction of radical pairs are obtained from future works, this mechanism can also be used to analyze the effect of magnetic particles, which exist universally in migrant birds [41], on the radical-pair product yield.

## III. CALCULATION MODEL

To analyze the resultant magnetic field produced by the interaction between the magnetic particles and the geomagnetic field, the dendrites in the upper beak that contain magnetic particles were simulated as a cylinder. The cylinder length was designated as $l$=300 μm, and the diameter was designated as $D$=10 μm, by evaluating the size of the magnetic particles and dendrites [10], as shown in Fig. 1(a).

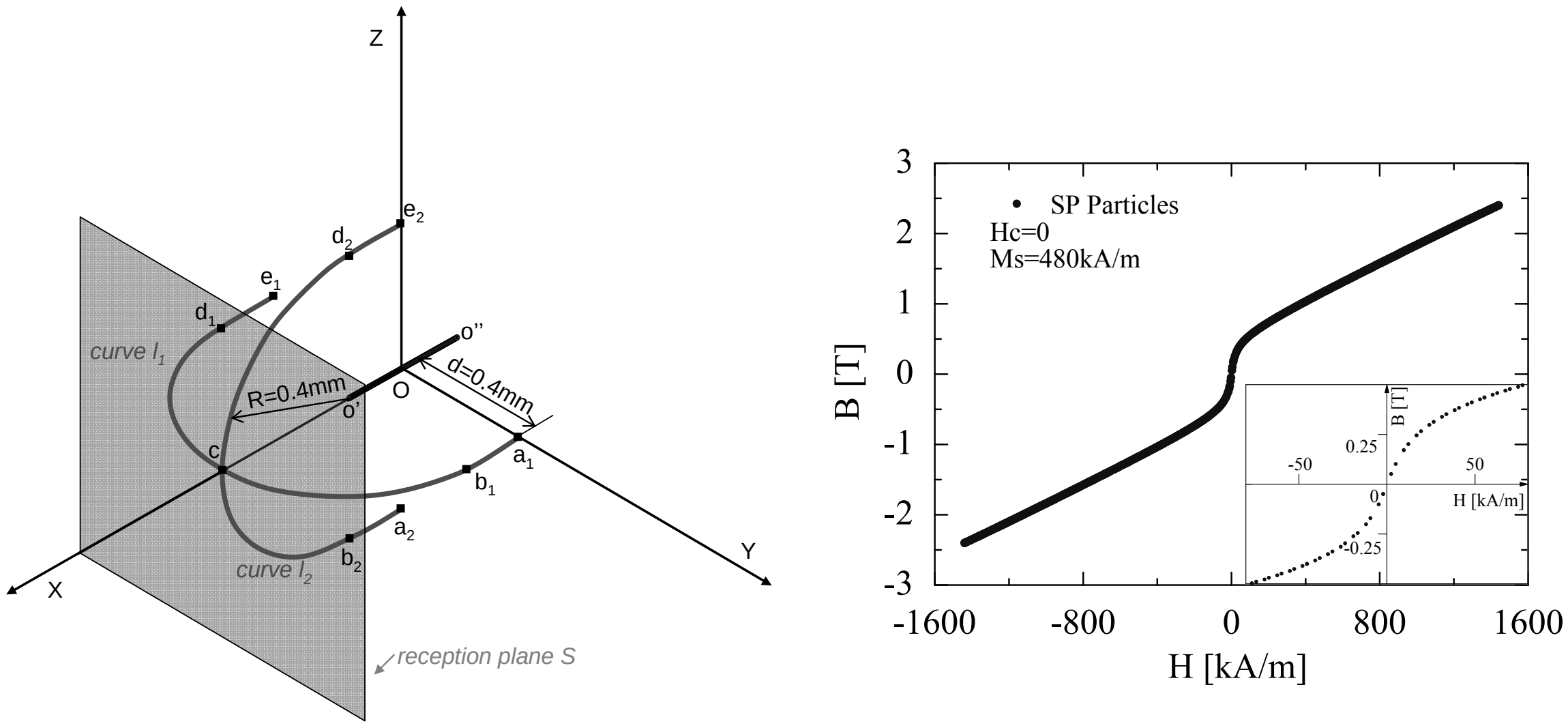

(a) (b)

**Fig. 1 Magnetic field calculation.**

(a) Three-dimensional model for magnetic particles. The cylinder is used to simulate the magnetic particles in birds. The centre of the cylinder is the origin of the coordinate system for calculation. (b) Experimental *B*–*H* curve of the SP nanoparticles measured by VSM.

The *x*-axis is defined as the central axis of the cylinder, the *y*-axis is perpendicular to the central axis of the cylinder, and the *z*-axis is perpendicular to the *xy* plane and parallel to the left and right surfaces of the cylinder. This rough assumption is not suitable for quantitative calculation because of insufficient biophysical experimental data, but it is sufficient for qualitative analysis of the effects of the resultant magnetic field on the radical pair product yield.

To investigate the resultant magnetic field distribution, curve $l_1$ in the $xy$ plane and curve $l_2$ in the $xz$ plane, shown in Fig. 1(a), were chosen to calculate the amplitude of the resultant magnetic field at different positions when the applied magnetic field changes. $a_1$–$b_1$ and $a_2$–$b_2$ and $d_1$–$e_1$ and $d_2$–$e_2$ are line segments parallel to the $x$-axis; the distance from $o'$ (centerpoint of the left surface of the cylinder) to the line is ±0.4 mm. $b_1$–$d_1$ and $b_2$–$d_2$ are arcs with a radius of 0.4 mm and an angle of 180°, whose centerpoint is $o'$.

For convenience of calculation, the bird's body was assumed to be parallel to the $x$-axis of the reference coordinate system. Therefore, when the bird changes its angle of flight attitude, the direction of the geomagnetic field (not the bird's body) would rotate in the reference frame. Magnetic field $\boldsymbol{B}_{ext}$, whose included angle $\theta$ with the $x$-axis changes from -180° to 180°, was applied as a boundary condition to simulate the geomagnetic field. Assuming that magnetic field $\boldsymbol{B}_{ext}$ is in the $xz$ plane, the $x$ component of the external magnetic field is $B_{ext_x}=B_0\cos\theta$, the $y$ component is $B_{ext_y}=0$, and the $z$-component is $B_{ext_z}=B_0\sin\theta$, where $B_0=|\boldsymbol{B}_{ext}|=50$ μT. With regard to the magnetic material properties, two kinds of magnetic materials were proposed to be involved in the magnetoreception process.

a) SP nanoparticles: This proposal was based on the findings of the SP nanoparticles in birds using transmission electron microscope (TEM) imaging in bright and dark-field mode and using small-area electron diffraction (SAED) [4]. The SP particles, whose coercivity is close to zero, were considered as an amplifier of the geomagnetic field during the magnetoreception process. Considering the lack of measurement results of the specific properties of magnetic particles in birds, the experimental results of the SP nanoparticle (diameter: 10 nm, produced by Ferrotec Corporation, Japan) was used to determine the magnetic properties. The $B$–$H$ curve shown in Fig. 1(b) was measured using a vibrating sample magnetometer (VSM). The coercivity of the SP material for the magnetic field calculation obtained from the $B$–$H$ curve is zero, and the relative permeability is 16. The measurement sample of the SP nanoparticle was a magnetic fluid and not pure SP nanoparticles. To obtain the magnetic properties of the SP materials, the curve was amplified using the saturation magnetization of $Fe_3O_4$ ($M_s$=480 kA/m).

b) Ferromagnetic materials: This proposal was based on the findings of two different types of iron compounds in the upper beak skin of adult homing pigeons using different light and electron microscopic methods combined with X-ray analysis. These two iron compounds were identified as two ferromagnetic materials: magnetite and maghemite [14], which closely resemble the real iron-containing structure in birds. However, the specific magnetic properties of the two types of materials have not been identified by measurements yet. In comparison with the SP materials, the coercivity of the ferromagnetic materials for magnetic field calculation was set to 17 kA/m, and the relative permeability was set to 16. In addition, the residual magnetization direction of the magnetic cylinder was considered to be along the $x$-axis (i.e., central axis of the cylinder). These parameters were almost the same as those used in the magnetic particle simulation model of Solov'yov et al. [15].

The resultant magnetic fields corresponding to these two kinds of magnetic materials were calculated and used to analyze the product yield of the radical pair. As this analysis was a large-scale qualitative simulation, the assumption for the two magnetic materials did not consider the special characteristics of the magnetic particles in the nanometer scale.

The isotropic radical-pair magnetic field effect was used to calculate the singlet state product yield of the reactions under different magnetic field strengths using the method of Timmel et al. [42]. This method was considered in the case of a radical pair with a single spin-1/2 nucleus (e.g., a proton), which ignored radical-pair diffusion and allowed singlet/triplet pairs to disappear with the first-order kinetics with rate constants $k_S=k_T=k$ [43].

The formula for calculating the singlet state product yield $\Phi_S$ is shown in Eq. (1), where $\omega$ is the Larmor frequency ($\omega=\gamma B$) and $a$ is the isotropic hyperfine coupling constant.

$$\begin{aligned}\Phi_s = {} & \frac{3}{8}+\frac{1}{8}\cdot\frac{\omega^2}{\Omega^2}+\frac{1}{8}\cdot\frac{a^2}{\Omega^2}\cdot f(\Omega) \\ & +\frac{1}{8}\cdot(1-\frac{\omega}{\Omega})\cdot f(\frac{1}{2}a+\frac{1}{2}\omega+\frac{1}{2}\Omega) \\ & +\frac{1}{8}\cdot(1-\frac{\omega}{\Omega})\cdot f(\frac{1}{2}a-\frac{1}{2}\omega-\frac{1}{2}\Omega) \\ & +\frac{1}{8}\cdot(1+\frac{\omega}{\Omega})\cdot f(\frac{1}{2}a-\frac{1}{2}\omega+\frac{1}{2}\Omega) \\ & +\frac{1}{8}\cdot(1+\frac{\omega}{\Omega})\cdot f(\frac{1}{2}a+\frac{1}{2}\omega-\frac{1}{2}\Omega)\end{aligned} \quad (1)$$

In addition, $f(x)$ is the Lorentzian function, where

$$f(x)=\frac{k^2}{k^2+x^2} \quad (2)$$

$$\Omega=(a^2+\omega^2)^{1/2} \; . \quad (3)$$

## Ⅳ. CALCULATION RESULTS

### A. Magnetic field distribution

The relationship among the amplitude of the resultant magnetic field $|B|$, the included angle $\theta$ between the applied magnetic field and the central axis of the magnetic cylinder, and length $L_i$ in curves $l_1$ and $l_2$, calculated in the cylindrical magnet of SP and the ferromagnet, is shown on the 3-D diagram in Fig. 2. $L_i$ is defined as the length away from point $a_i$ in curve $l_i$ ($i$=1, 2). When $\theta$ is 0°, the $|B|$ distribution corresponding to distance $L_1$ in curve $l_1$ is the same as that of $L_2$ in curve $l_2$ because of axial symmetry, as shown in Fig. 3(a) and (b). The relationship between $|B|$ and $d_x$ (distance from $o'$ to the calculation point at the $x$-axis) is shown in Fig. 3(c).

To investigate the effect of the change in the magnetic field amplitude ($\Delta|B|=|B|_{max}-|B|_{min}$, where $|B|_{max}$ is the maximum value of $|B|$ and $|B|_{min}$ is the minimum value of $|B|$) on the product yield of a radical pair, $|B|$ of centerpoint $c$ in curves $l_1$ and $l_2$ (i.e., $L_i\approx800$ μm) in the SP and ferromagnetic materials was calculated. This process also compared the difference between the two materials. The relationships between $|B|$ and $\theta$ in the SP and ferromagnetic materials with $d_x$=0.4 mm are shown in Fig. 4(a) and (b), respectively ($\Delta|B|_{SP}\approx0.03$ μT and $\Delta|B|_{Ferromagnet}\approx24$ μT).

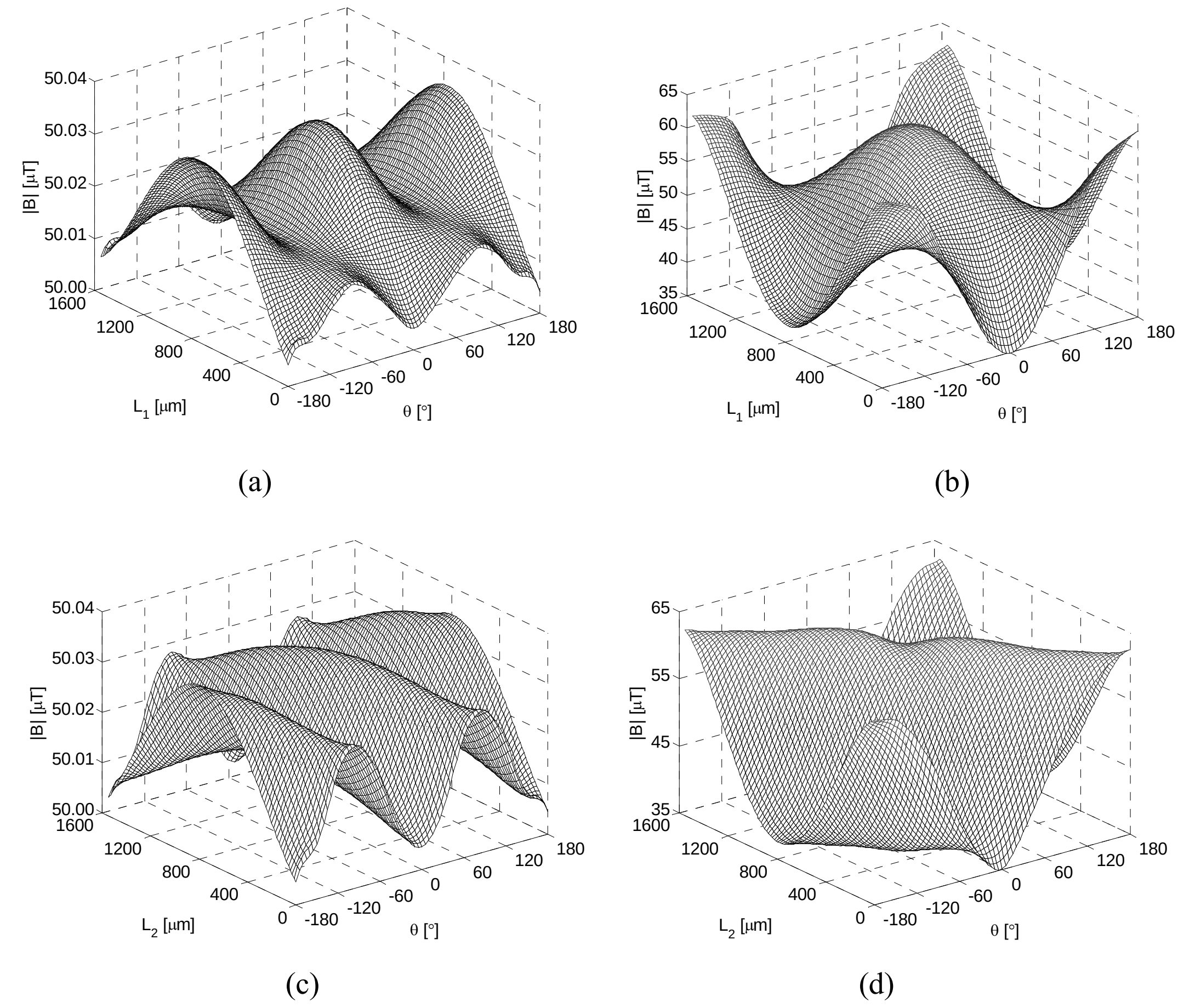

**Fig 2. Resultant magnetic field |*B*| distribution when $\theta$ varies from -180° to 180°.**

(a) Relationship among |*B*|, $\theta$, and the position of the observed point in curve $l_1$ in the SP material. (b) Relationship among |*B*|, $\theta$, and the position of the observed point in curve $l_1$ in the ferromagnetic material. (c) Relationship among |*B*|, $\theta$, and the position of the observed point in curve $l_2$ in the SP material. (d) Relationship among |*B*|, $\theta$, and the position of the observed point in curve $l_2$ in the ferromagnetic material.

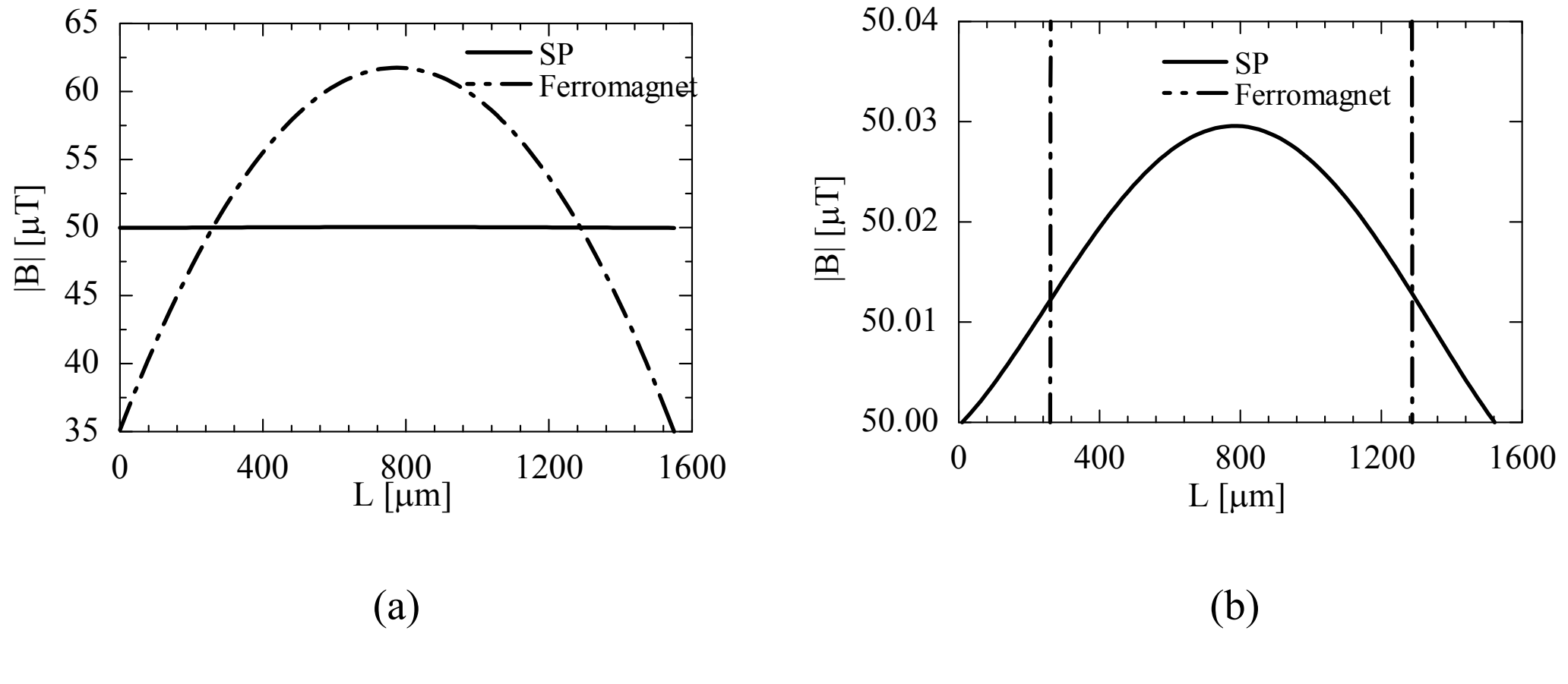

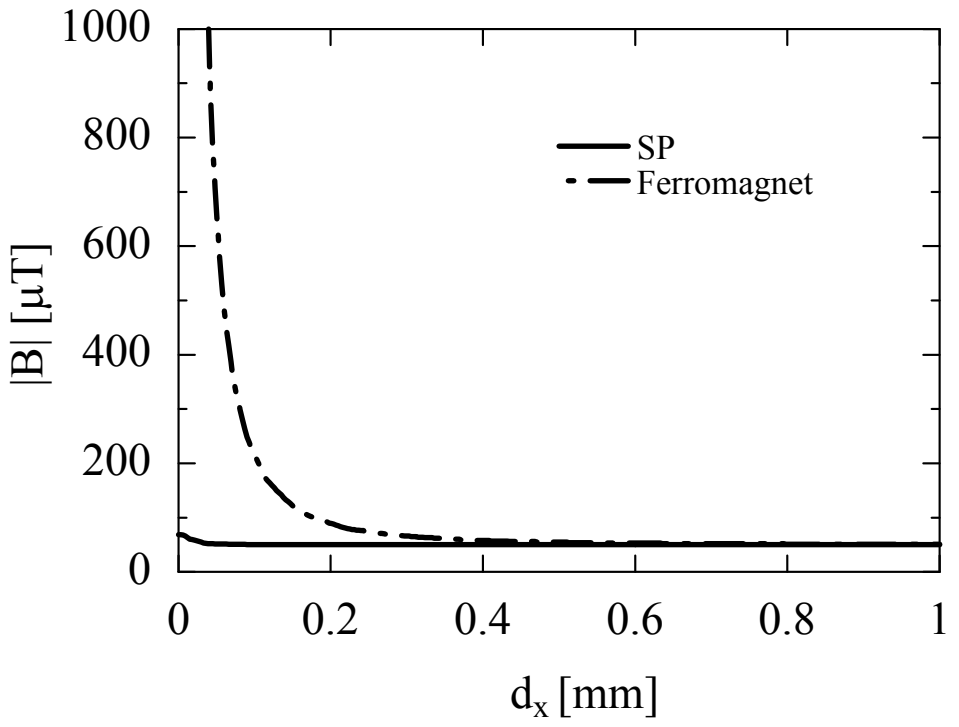

(c)

**Fig 3. Distribution of |*B*| at different positions when *θ*=0°.**

(a) Distribution of |*B*| along curve $l_1$ or $l_2$. (b) Partial enlargement of Fig. 4(a). (c) Distribution of |*B*| along the *x*-axis.

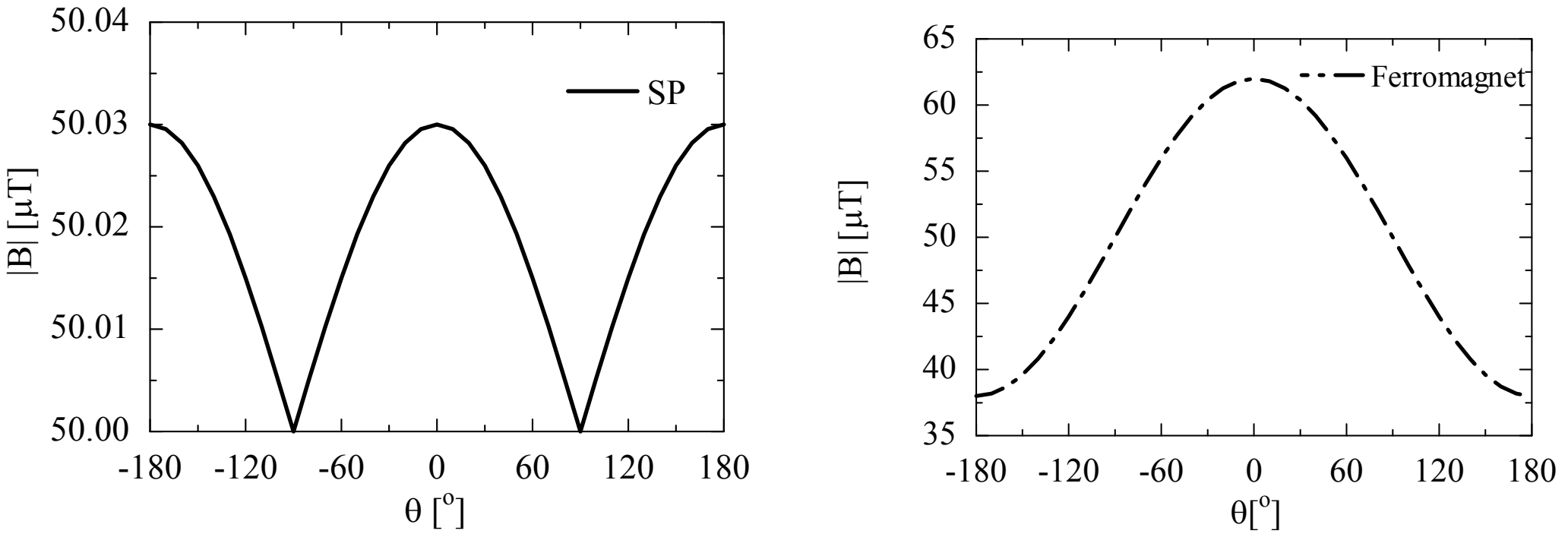

(a) (b)

**Fig 4. Relationship between |*B*| and *θ* at $d_x$=0.4 mm.**

(a) In the SP material. (b) In the ferromagnetic material.

B. Singlet state product yield corresponding to the magnetic field

The singlet state product yield $\Phi_S$ was calculated according to the method mentioned in Section III. The relationship between the singlet state product yield $\Phi_S$ and |*B*| is shown in Fig. 5(a). To facilitate discussion, the horizontal axis considered in the figure was |*B*| and not |*B*|/*a*, although the latter is more commonly used in other papers. In addition, based on the experimental results in a protein environment using flash photolysis [44], the decay rate *k* was set as 1 $\mu s^{-1}$. According to the calculation formula, the different values of hyperfine coupling constant *a* influenced the $\Phi_S$ range corresponding to the change in the magnetic field direction angle *θ* from -180° to 180° located in different parts of the curve, as shown in Fig. 5(b) and (c). If *a* is much higher than the Larmor frequency of |*B*|, the $\Phi_S$ range corresponding to the magnetic field direction angle *θ* monotonically decreases, as shown by

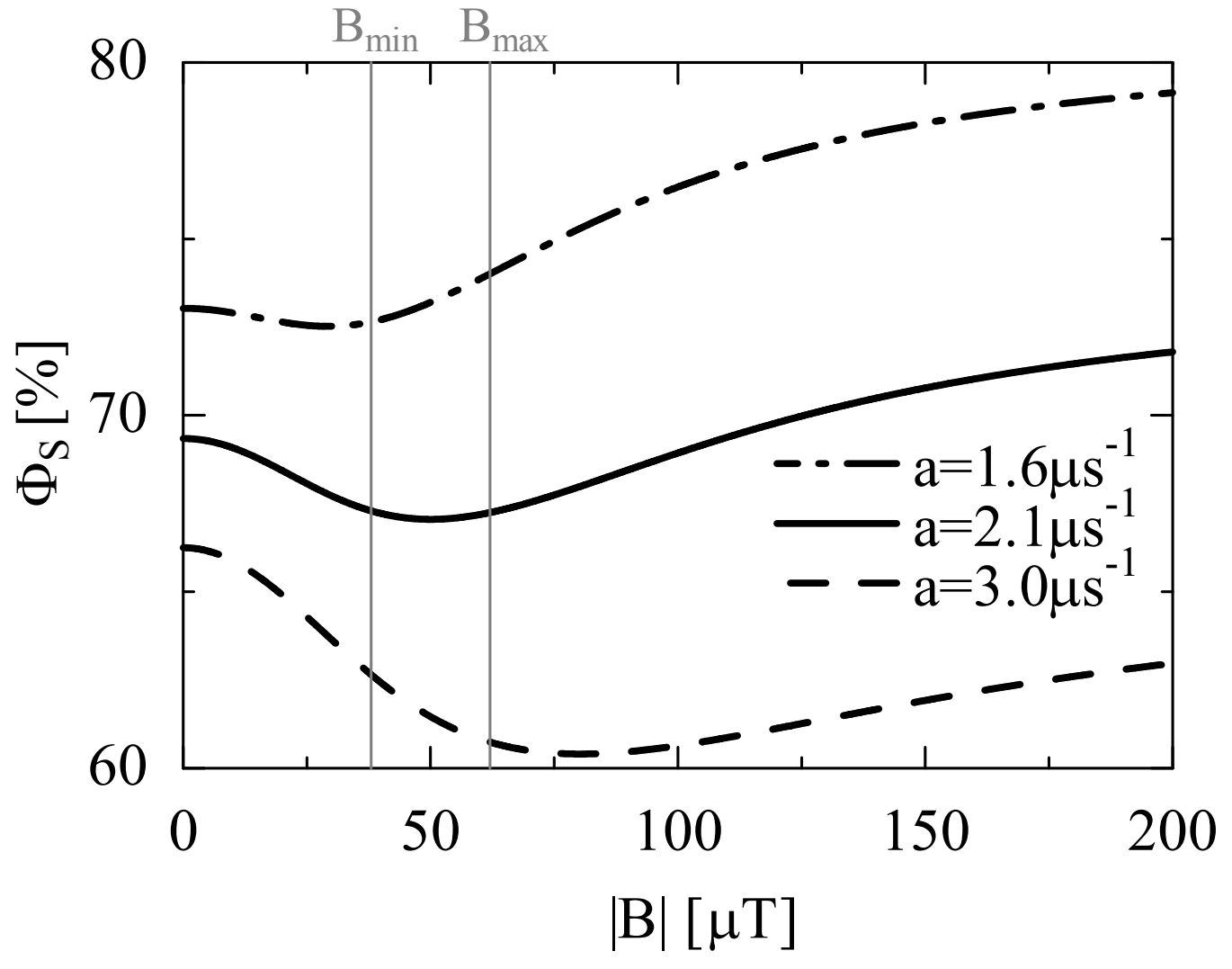

(a)

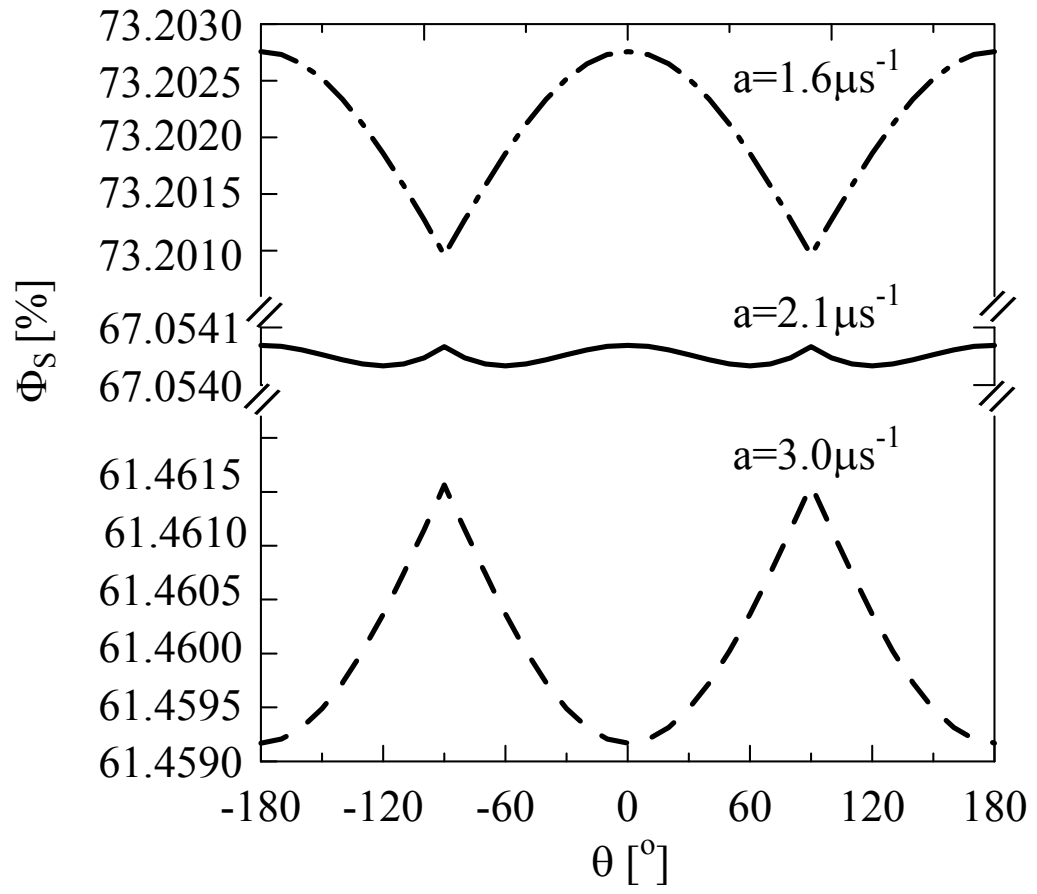

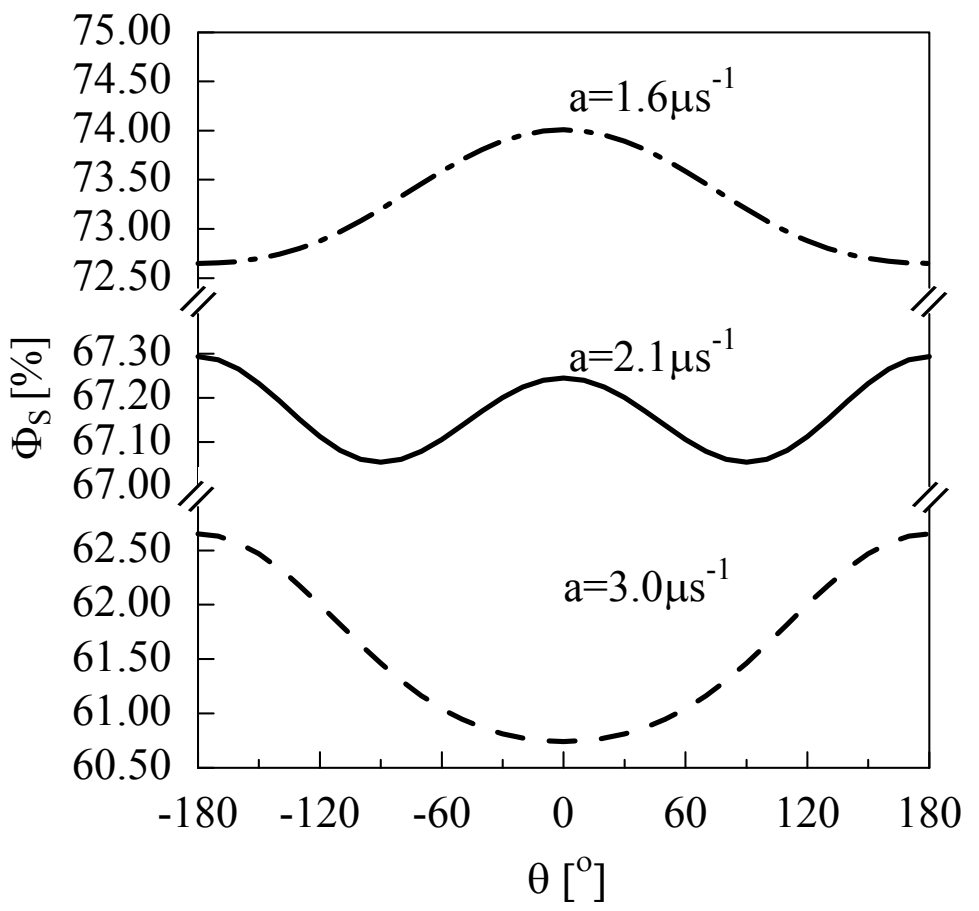

(b) (c)

**Fig 5. Variation in the singlet state product yield $\Phi_S$ of the radical pairs.**

(a) Relationship between $\Phi_S$ and $B/a$. (b) Relationship between $\Phi_S$ and $\theta$ in the SP material. (c) Relationship between $\Phi_S$ and $\theta$ in the ferromagnetic material.

the dashed line ($a$=3.0 $\mu s^{-1}$) in Fig. 5(a). The relationships between $\Phi_S$ and $\theta$ in the SP and the ferromagnetic materials are shown by the dashed lines in Fig. 5(b) and (c), respectively. If $a$ is much lower than the Larmor frequency of $|B|$, the $\Phi_S$ range corresponding to the magnetic field direction angle $\theta$ monotonically increases, as shown by the dot–dashed line ($a$=1.6 $\mu s^{-1}$) in Fig. 5(a). The relationships between $\Phi_S$ and $\theta$ in the SP and ferromagnetic materials are shown by the dot–dashed lines in Fig. 5(b) and (c), respectively. If the value of $a$ is appropriate, the $\Phi_S$ range can also include the turning point, as shown by the solid line ($a$=2.1 $\mu s^{-1}$) in Fig. 5(a). Under this condition, the relationship between $\Phi_S$ and $\theta$ in the SP and ferromagnetic materials are shown by the solid lines in Fig. 5(b) and (c), respectively.

## C. Modulation patterns corresponding to the geomagnetic information

The results presented earlier only indicated the variation of $\Phi_S$ corresponding to the magnetic field direction at a specific point. Generally, the sensory receptors of organisms are assumed as ordered structures; thus, a reception plane is chosen to investigate the singlet state product yield distribution of the ferromagnetic material in this section. Assuming that the residual magnetization direction of the magnetic cylinder is along the bird's body axis, the sensory receptors are arranged in reception plane *S* (area: 0.5 mm×0.5 mm) parallel to the *yz* plane. The distance of reception plane *S* along the positive direction of the *x*-axis from *o*' is 0.4 mm, as shown by the gray quadrangle in Fig. 1(a). According to the proposed model, when the applied magnetic field rotates parallel to the *xy* and *xz* planes (the included angle $\theta$ with the *x*-axis changes from 0° to 180°), the $\Phi_S$ distribution corresponding to $|B|$ is calculated as shown in Fig. 6.

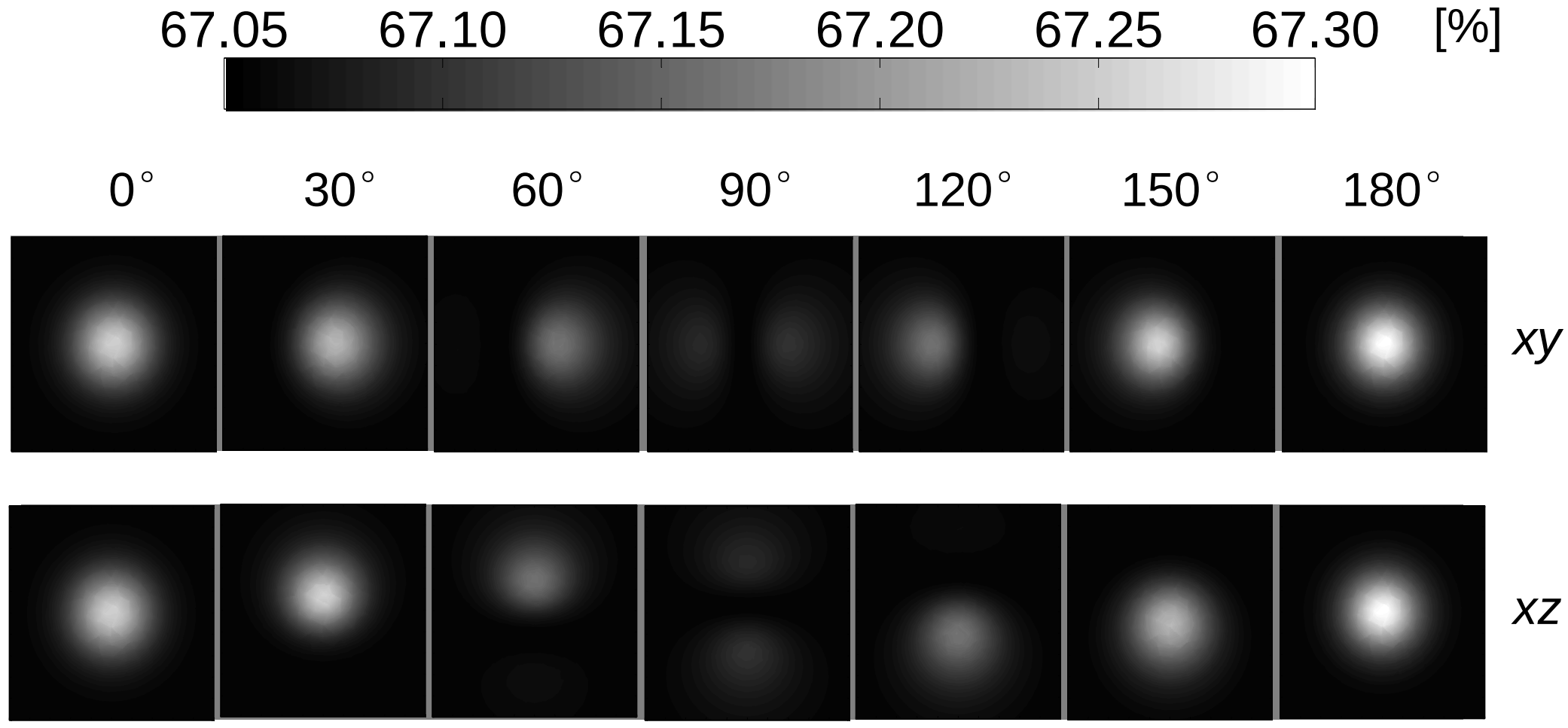

Fig. 6. Modulation patterns of $\Phi_S$ through the magnetic field parallel to the *xy* and *xz* planes in different directions at angles 0, 30, 60, 90, 120, 150, and 180°.

## Ⅴ. DISCUSSION

Compared with the research conducted by Cohen and Cai, the difference in the combined mechanisms in this paper is that we focused on the amplitude of the resultant magnetic field around the magnetic particles instead of the magnetic field gradient. This process means that the changes in the singlet state product yields correspond directly to the magnetic field at the same spatial position, and the difference in the local magnetic fields between the two radicals of each radical pair was ignored. This result can be attributed to the difference in the local magnetic fields of the two radicals, which is much smaller than the changes in the resultant magnetic field amplitude $\Delta|B|$ during the geomagnetic field rotation in the proposed mechanism. For example, when the distance from the magnetic particles $d_x$=0.4 mm, $\Delta|B|_{SP}\approx 0.03$ μT, and $\Delta|B|_{Ferromagnet}\approx 24$ μT. If the distance between two radicals is 3.5 nm (the same as the value used in Cai's calculation [39]), the variation in the local magnetic fields of the two radicals is approximately $5.29\times10^{-5}$ μT. In Cai's calculation, it was approximately 4 mT, considerably larger than our calculation result, possibly because Cai assumed that the acceptor of the radical pairs was 35 nm away from the magnetic particles. Therefore, the proposed model appears more reasonable in cases when the distance between the radical pairs and the magnetic particles is relatively farther.

Based on the magnetic field distribution calculation results, $|B|$ was found to change cyclically with included angle $\theta$, and the period of variation was 180° in the SP materials and 360° in the ferromagnetic materials. When $\theta$ was 0°, $|B|$ of centerpoint $c$ in the two kinds of materials both reached the maximum. The $|B|_{SP}$ variation was similar to the function $f(\theta)=|\sin(\theta-90°)|+offset$. The $|B|_{Ferromagnet}$ variation at centerpoint $c$ was similar to cosine waves with an offset because ferromagnetic materials have inherent remanence, whereas SP materials do not have any.

In the calculation, $\Delta|B|_{SP}$ is very small because $|B|$ decays exponentially with $d_x$; if $d_x$ is smaller, then $\Delta|B|_{SP}$ would become higher. For example, if $d_x$=0.1 mm, then $\Delta|B|_{SP}\approx0.3$ μT, and if $d_x$=10 μm, then $\Delta|B|_{SP}\approx20$ μT. In all cases, the variation trends of $|B|$ in the SP material with different $d_x$'s are almost similar. Therefore, the calculation result with $d_x$=0.4 mm can still be used for qualitative analysis.

At the same spatial position as in point $c$ in Fig. 1(a), $|B|_{Ferromagnet}$ is much larger than $|B|_{SP}$. The change in the singlet state product yields $\Delta\Phi_S = \Phi_{Smax} - \Phi_{Smin}$ of the ferromagnetic material is also much larger than $\Phi_S$ of the SP material. When $d_x$=0.4 mm ($\Delta|B|_{SP}\approx0.03$ μT and $\Delta|B|_{Ferromagnet}\approx24$ μT), the decay rate is $k$=1 μs$^{-1}$, and the values of hyperfine coupling constant $a$ are 1.6, 2.1, and 3.0 μs$^{-1}$. The calculated $\Delta\Phi_S$ of the SP materials are approximately $1.8\times10^{-5}$, $3.6\times10^{-7}$, and $2.4\times10^{-5}$, and $\Delta\Phi_S$ of the ferromagnetic materials are approximately 1.36%, 0.24%, and 1.91%, respectively. The changes in product yields $\Phi_S$ calculated at present are relatively small. If $d_x$ decreases slightly (but not below micrometer level), then $\Delta\Phi_S$ would become more significant. For example, when $d_x$=0.1 mm, $\Delta\Phi_S$ of the SP material corresponding to the same values of hyperfine coupling constant $a$ of 1.6, 2.1, and 3.0 μs$^{-1}$ would be approximately $1.8\times10^{-4}$, $1.2\times10^{-6}$, and $2.4\times10^{-4}$, respectively ($\Delta|B|_{SP}\approx0.3$ μT). $\Delta\Phi_S$ of the ferromagnetic materials are approximately 6.56%, 3.03%, and 2.59%, respectively ($\Delta|B|_{Ferromagnet}\approx250$ μT), which would mean that the reception radical pairs are probably closer to the magnetic particles and not located in the retina of the birds, according to the proposed mechanism.

We can reasonably assume that nature has optimized the properties through evolution to provide maximum results; therefore, these calculation results can explain why biophysical experiments showed that the magnetic particles in organisms are more similar to ferromagnetic (and not SP) materials [14]. The former can provide more significant changes in product yield $\Phi_S$, which made magnetoreception more feasible.

Moreover, when the range of singlet state product yield $\Phi_S$ includes a turning point in the ferromagnetic material or the $\Phi_S$ range monotonically decrease or increase in the SP material, the values of $\Phi_S$ corresponding to the magnetic field direction angle $\theta$=0° and $\theta$=180° both attained the maxima (the values are almost equal), indicating that the magnetic compass is an inclination compass. This conclusion is in agreement with the experimental results obtained from a variety of research activities on migratory birds [45], where the formula for calculating the singlet state product yield $\Phi_S$ was based on one-proton radical pair. In fact, when the radical pairs are more complex, the calculation formula will also become more complex. However, the curve of $\Phi_S$ corresponding to $|B|$ also has an extreme value; therefore, the proposed model is still suitable.

In addition, because the experimental results showed that the properties of the magnetite crystals in birds are more similar to ferromagnetic materials, SD magnetic materials were not considered in the model calculation. In fact, SD magnetic materials will only make the magnetic field around the magnetic particles higher and, hence, induce more significant changes in the product yield, but the variation trend of $\Phi_S$ calculated with SD materials is still similar to that of the ferromagnetic materials.

Based on the response pattern of the singlet state product yield $\Phi_S$, the geomagnetic information was found to be represented by the distribution of $\Phi_S$ in the reception plane. The position of high $\Phi_S$ is located at the center position of the reception plane because the body axis is parallel to the geomagnetic field. If the included angle between the body axis and the geomagnetic field is changed in the horizontal or vertical plane, the position of the high $\Phi_S$ will move correspondingly along the horizontal or vertical direction. The pattern for a bird flying anti-parallel to the magnetic field direction (180°) is similar to the pattern for parallel orientation (0°), which also demonstrates that the magnetic compass of the birds is intrinsically an inclination compass. Birds can use modulation patterns to adjust the flight direction.

The previous discussions ignored the rotation of the bird itself around the body axis during the flight, which often happens during the bird's attitude adjustment process. However, the bird can receive the rotation information from multiple magnetic particle arrangements or from other means (such as the gravity information response) to assist in navigation. Furthermore, the modulation patterns are very similar to the results obtained by radical-pair-based mechanism only [7, 8], but the biophysical mechanisms are completely different.

## Ⅵ. CONCLUSION

This paper has proposed a biophysical mechanism for magnetoreception in birds, which combined the magnetite- and radical-pair-based mechanisms to provide a more reasonable explanation of avian magnetoreception. The amplitude of the resultant magnetic field produced by the interaction between the magnetic particles and the geomagnetic field was analyzed using two materials with different magnetic properties. The relationship between the direction of the geomagnetic field and the radical-pair product yield was investigated. The calculation results showed that the period of variation for singlet state products is approximately 180° when the magnetic particles are ferromagnetic materials, which could explain the experimental results of avian magnetic compass research. The singlet state products through the magnetic field in a reception plane were investigated. Their modulation patterns can provide geomagnetic information in birds. The model calculation results showed that the proposed model, which deals with the magnetic particles' role in changing the direction information of the geomagnetic field to amplitude of the resultant magnetic field, can explain how birds use geomagnetic field for navigation better than using either magnetite- or radical-pair-based mechanism only.

## ACKNOWLEDGMENTS

The authors thank Dr. Chunxiao Xu for valuable discussions. This work was supported by the State Key Program of National Natural Science of China (51037006). This work was supported by the State Key Development Program for Basic Research of China (2011CB503702).